\documentclass[aip,
 amsmath,amssymb,
 reprint,%
]{revtex4-1}

\usepackage{graphicx}
\usepackage{dcolumn}
\usepackage{bm}

\usepackage[utf8]{inputenc}
\usepackage[T1]{fontenc}
\usepackage{mathptmx}

\begin{document}

\preprint{AIP/123-QED}

\title{Diode effect in the lateral spin valve}

\author{E.A. Karashtin}
 \email{eugenk@ipmras.ru.}
 \altaffiliation[Also at ]{Lobachevsky University, Gagarin ave., 23, Nizhniy Novgorod, 603950, Russia}
\author{D.A. Tatarskiy}
 \email{tatarsky@ipmras.ru.}
 \altaffiliation[Also at ]{Lobachevsky University, Gagarin ave., 23, Nizhniy Novgorod, 603950, Russia}
\affiliation{Institute for physics of microstructure of RAS, Nizhniy Novgorod, 603950, GSP-105, Russia}

\date{\today}

\begin{abstract}

We propose a possible experimental setup for nonreciprocal electron transport in a lateral spin valve due to noncoplanar distribution of magnetic moment (and field) in the system. Some metals (Al, Cu) and semiconductors (GaAs, InSb etc.) demonstrate spin accumulation even at room temperatures due to large spin flip lengths (0.3--2.0 $\mu m$). The Hanle precession was observed in lateral spin valves based on such materials and two parallel magnetic electrodes. We provide theoretical estimations which show that the nonreciprocal effect in a configuration with non-parallel magnetic electrodes is of the same order of value as the effect that arise due to Hanle precession. This makes it possible to observe the nonreciprocal electron transport as a manifestation of the noncommutativity of spin-1/2 algebra for them in the proposed system. Such property of noncoplanar magnetic system has a potential of application in a noncoplanar type of spintronic devices.

\end{abstract}

\maketitle

\section{\label{Intro}Introduction}
Spatial noncoplanar magnetic field distributions can demonstrate unusual transport effect for spin-$\frac{1}{2}$ particles: persistent electric current~\cite{Tatara,Balatsky}, rectification effects for neutrons~\cite{Tatarskiy,UdalovHelNeutrons} and electrons~\cite{UdalovDiode,Karashtin,UdalovSpin}, neutrons ``skew'' scattering~\cite{UdalovHall,UdalovSkew} and topological Hall effect for electrons~\cite{Skyrm_THE,Averkiev}. The detection of persistent current or rectification effect in typical ferromagnets is very complicated due to small value of these effects and the intensive spin-flip scattering of carriers. Thus one should consider systems where the spin-flip scattering is low. For example, thermal neutrons have a spin-flip length about 10 cm on the air in the Earth magnetic field. This makes it possible to observe the diode effect for neutrons in transmission through two magnetic mirrors in an external field~\cite{Tatarskiy2}.

Aluminium and copper have very large spin-flip length~\cite{Jedema} for conductive electrons. Thus the polarized electron transport can be observed in a lateral spin valve based on such materials~\cite{Jedema2,Silsbee}. The $A_3 B_5$ semiconductors are very promising materials for lateral spin-dependent transport ~\cite{SemiSF,GaAsSF,InSbSF1}. E.g. InSb has a very large spin-flip length (several $\mu m$) at room temperature~\cite{InSbSF2}. This allows to develop an artificial system for observation of nonreciprocal electron transport in case of noncoplanar spatial distribution of a magnetic field.

In this paper we consider a lateral spin valve~\cite{Jedema} in which the magnetizations of the ferromagnetic electrodes have arbitrary angles with respect to the spin flow direction. The external field is applied perpendicularly to the electrode magnetizations~(Fig.~\ref{Fig1}). Thus the magnetizaitons of the electrodes and the applied magnetic field form a noncoplanar triplet. The diode effect means the resistivity should depend on the direction of charge carriers injection. This system is analogous to the nonreciprocal cell for thermal neutrons~\cite{Tatarskiy2}. The rectification effect was previously calculated in a similar F/I/N/I/F system~\cite{UdalovSpin}. Here we consider one-dimensional channel for charge carriers. We show the nonreciprocal diode effect is of the same order as a typical Hanle precession in a coplanar lateral spin valve and therefore may be observed in such system. Carrying out such experiment would manifest first step towards spintronic devices~\cite{Zutic,Spin1,Spin2,Spin3} that utilize special and non-trivial properties of noncoplanar magnetization distribution.

\section{\label{Sec2}Theoretical approach}
\begin{figure}[t]
\includegraphics[width=3.2in, keepaspectratio=true]{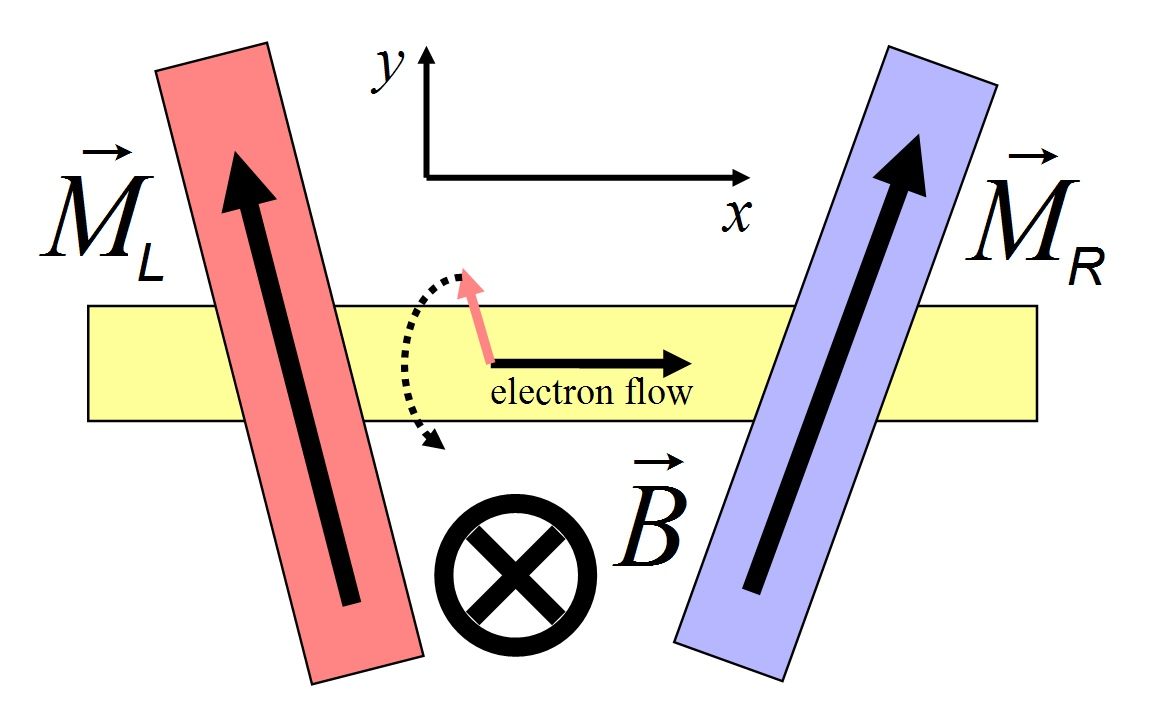}
\caption{\label{Fig1} (Color online) Geometry of noncoplanar lateral spin valve.}
\end{figure}

We consider a system depicted on Figure~\ref{Fig1}.

It consists of two ferromagnets connected electrically by a non-magnetic one-dimensional channel through thin insulating interlayer. The ferromagnets are magnetized noncollinearly. An external magnetic field is applied to the system in the direction perpendicular to both magnetizations of ferromagnets. The magnetic field is supposed to be small enough to keep magnetizations of ferromagnets almost unchanged. Besides, it is supposed that spin diffusion length in a conductive channel is bigger than its length (between two ferromagnets). The electric current passes from one ferromagnet to another. We calculate its dependence on the applied voltage and thus find the resistance of the system.

In the considered system the charge transport is determined by three processes. First one is the injection of spin-polarized carriers from one ferromagnet (spin polarizer) to the one-dimensional conducting channel. Second one is the transport of injected non-equilibrium charge to the second ferromagnet (spin analyzer). Finally, third process is tunnelling of charge carriers from the conducting channel to the spin analyzer. In order to describe transport of spin-polarized electrons in the channel we use the Boltzmann equations with a four-component distribution function
\begin{equation} \label{Eq2_1}
\hat{f} =
\begin{pmatrix}
f_{++}& f_{+-}\\
f_{-+}& f_{--}
\end{pmatrix}
\end{equation}
following Hernando et al.~\cite{Brataas1}, where indices ``+'',``-'' denote spins states along the quantization axis $z$. The tunnelling of electrons at both boundaries is described by boundary conditions that provide conservation of electron mass and spin flux \cite{Brataas2}.

\subsection{Equations for electrons in a channel}
The distribution function of conductivity electrons may be re-written as
\begin{equation} \label{Eq2_2}
\hat{f} = n\left(x,v_x\right) \sigma_0  + \mathbf{m}\left(x,v_x\right) \cdot \mathbf{\sigma},
\end{equation}
where $\sigma_0$ is the 2$\times$2 identity matrix, $\mathbf{\sigma}$ is the vector of Pauli matrices, $x$ and $v_x$ are the Cartesian coordinate and speed along the channel respectively. Obviously, $n$ in (\ref{Eq2_2}) stands for the local electron concentration distribution, while $\mathbf{m}$ is their magnetic moment distribution function.

If the spin polarizer and analyzer are divided from the intermediate channel by the insulating barriers we may neglect the electric field inside the conductive channel. Only the external magnetic field needs to be taken into account. Another important assumption is the tau-approximation for scattering integrals. Besides, we suppose that there are three times: ultrafast momentum relaxation time $\tau_p$, which is referred only for elastic scattering processes, relatively slow spin relaxation time $\tau_s$ and an even more slower energy relaxation time $\tau_e$. The Boltzmann equations in steady state take the following form
\begin{eqnarray} \label{Eq2_3}
&v_x& \frac{\partial n}{\partial x} = - \frac{n-\overline{n}}{\tau_p} -\frac{\overline{n}-n_0}{\tau_e}, \\ \label{Eq2_4}
&v_x& \frac{\partial \mathbf{m}}{\partial x} + \gamma_e \left[\mathbf{B} \times \mathbf{m}\right] =  - \frac{\mathbf{m}-\overline{\mathbf{m}}}{\tau_p} -\frac{\overline{\mathbf{m}}-\mathbf{m}_0}{\tau_s},
\end{eqnarray}
$\gamma_e$ is the electron gyromagnetic ratio, $\mathbf{B}$ is the applied magnetic field which is assumed to be perpendicular to the plane of the system. Here we introduce the averaged over momentum concentration and magnetization density $\overline{n}$ and $\overline{\mathbf{m}}$, respectively. Besides, ``equilibrium'' values $n_0$ and $\mathbf{m}_0$ are used. These values are the average of $\overline{n}$ and $\overline{\mathbf{m}}$ over energy, respectively. Since we suppose that the momentum relaxation time $\tau_p$ is much smaller than all other characteristic times we may obtain separate equations for $\overline{n}, \overline{\mathbf{m}}$ and $\delta n = n - \overline{n}, \delta \mathbf{m} = \mathbf{m} - \overline{\mathbf{m}}$
\begin{eqnarray} \label{Eq2_5}
&v& \frac{\partial \overline{n}}{\partial x} = - \frac{\delta n}{\tau_p}, \\ \label{Eq2_6}
&v& \frac{\partial \delta n}{\partial x} = -\frac{\overline{n}-n_0}{\tau_e}, \\ \label{Eq2_7}
&v& \frac{\partial \overline{\mathbf{m}}}{\partial x} + \gamma_e \left[\mathbf{B} \times \delta \mathbf{m}\right] =  - \frac{\delta \mathbf{m}}{\tau_p}, \\ \label{Eq2_8}
&v& \frac{\partial \delta \mathbf{m}}{\partial x} + \gamma_e \left[\mathbf{B} \times \overline{\mathbf{m}}\right] =  - \frac{\overline{\mathbf{m}}}{\tau_s}.
\end{eqnarray}
In these equations we suppose that the equilibrium magnetization is zero after relaxation, thus neglecting the electron spin polarization in the external magnetic field (this electron polarization may be neglected since it does not contribute to the nonreciprocal effect that appears in noncoplanar system). The velocity $v = \left|v_x\right|$ is defined by the energy as $v = \sqrt{\frac{2 \epsilon}{m_e}}$ ($\epsilon$ and $m_e$ are the electron energy and mass respectively). As it was mentioned earlier, the ``equilibrium'' electron density is $n_0 = \frac{\int{\overline{n}dv}}{\int{dv}}$.

The equations for electron density (\ref{Eq2_5}), (\ref{Eq2_6}) and for magnetic moment (\ref{Eq2_7}), (\ref{Eq2_8}) are solved separately. In order to obtain the electric current and spin current we need to find only the averaged over momentum parameters $\overline{n}, \overline{\mathbf{m}}$. We will further suppose that the magnetic field $\mathbf{B}$ is applied along the $z$-axis of the Cartesian coordinate system (see Figure~\ref{Fig1}). The equation for $\overline{n}$ takes the form
\begin{equation} \label{Eq2_10}
\frac{\partial^2 \overline{n}}{\partial x^2} = \xi_e^2 \left(\overline{n} - n_0\right),
\end{equation}
where $\xi_e = \left(\tau_e \tau_p v^2\right)^{-1/2}$ and $n_0$ is determined by $\overline{n}$ and therefore may depend on the coordinate.
We solve (\ref{Eq2_10}) for $\overline{n}$ supposing that the energy relaxation time is very big: $\xi_e d << 1$ for the Fermi velocity of electrons $v_F$, where $d$ is the length of the channel (see Figure~\ref{Fig1}). The solution then takes the form
\begin{equation} \label{Eq2_11}
\overline{n} = C_1 + C_2 \xi_e x + ...
\end{equation}
It is seen from the calculations below that the first two terms in (\ref{Eq2_11}) are enough. Thus the average electron density $n_0$ is defined by $C_1$ and $C_2$ but its dependence on the $x$-coordinate does not give correction to $\overline{n}$ in equation (\ref{Eq2_10}) up to the first order in $\xi_e x$.

General solution for $\overline{\mathbf{m}}$ has the form
\begin{eqnarray} \label{Eq2_20}
\overline{m}_x &=& W_1 e^{\left(a+ib\right)\xi_s x} + W_2 e^{\left(-a-ib\right)\xi_s  x} + c.c., \\ \label{Eq2_21}
\overline{m}_y &=& -i W_1 e^{\left(a+ib\right)\xi_s x} + i W_2 e^{\left(-a-ib\right)\xi_s x} + c.c., \\ \label{Eq2_22}
\overline{m}_z &=& A x + D,
\end{eqnarray}
where $W_1$ and $W_2$ are complex constants, $A$ and $D$ are real constants, $\xi_s = \left(\tau_s \tau_p v^2\right)^{-1/2}$, $a$ and $b$ are defined as
\begin{eqnarray} \label{Eq2_23}
a &=& \sqrt{\frac{\sqrt{1 + \beta^2 + \alpha^2 \beta^2} + 1 - \alpha \beta}{2}}, \\ \label{Eq2_24}
b &=& \mathbf{sign(B)} \sqrt{\frac{\sqrt{1 + \beta^2 + \alpha^2 \beta^2} - 1 + \alpha \beta}{2}}, \\ \label{Eq2_245}
\alpha &=& \omega_L \tau_p \equiv \gamma_e B \tau_p, \\ \label{Eq2_25}
\beta &=& \omega_L \tau_s \equiv \gamma_e B \tau_s.
\end{eqnarray}
Since the $z$-component of magnetization is not induced by the boundary conditions and we neglect the electron spin polarization due to $\mathbf{B}$, taking into account only precession of spin about it, the constants $A$ and $D$ are zero. It should be noted that although we do not suppose that the magnetic field $\mathbf{B}$ is small this supposition is implied because we do not take into account change of the magnetizations $\mathbf{M}_L$ and $\mathbf{M}_R$ by it.

It is convenient to write down all the conditions on parameters used to obtain the solution (\ref{Eq2_11}), (\ref{Eq2_20}), (\ref{Eq2_21}). These conditions are
\begin{eqnarray} \label{Eq2_30}
\tau_p &<<& \tau_e, \tau_s; \\ \label{Eq2_31}
v_F \sqrt{\tau_e \tau_p} &>>& d.
\end{eqnarray}
The condition (\ref{Eq2_30}) is realistic for Al conductors even at \textbf{low temperatures}~\cite{Jedema2,taup}. The condition (\ref{Eq2_31}) can be satisfied for relatively small conduction channel length (for estimations, see Section~\ref{Sec3}).

\subsection{Boundary conditions}
We suppose that there is no inelastic and spin-flip scattering inside thin barriers and at the boundaries. Therefore the boundary conditions are conservation of electron flow and spin current at the boundaries. The charge and spin current inside ferromagnets may be obtained using the Buttiker-Landauer formalism~\cite{Brataas2}. If the current flows from left ferromagnet located at $x = -d$ to right one located at $x = 0$ these currents take the form
\begin{eqnarray} \label{Eq2_90}
j^L_x &=& \frac{G}{2} \epsilon \left(1 - \overline{n}\left(-d\right)\right) \left(1 + P \mathbf{M}_L \cdot \overline{\mathbf{m}}\left(-d\right)\right), \\ \label{Eq2_91}
\mathbf{J}^{s \, L}_x &=& \frac{\hbar}{2} P' \left(1 - \overline{n}\left(-d\right)\right) \mathbf{M}_L
\end{eqnarray}
at the left bounday and
\begin{eqnarray} \label{Eq2_92}
j^R_x &=& \frac{G}{2} \epsilon \overline{n}\left(0\right) \left(1 - P \mathbf{M}_R \cdot \overline{\mathbf{m}}\left(0\right)\right), \\ \label{Eq2_93}
\mathbf{J}^{s \, R}_x &=& \frac{\hbar}{2} P' \overline{n}\left(0\right) \mathbf{M}_R
\end{eqnarray}
at the right boundary. Here $G$ is a constant defined by electron tunneling probability, $P$ is a constant that takes into account spin accumulation in the metal channel, $P'$ is a constant that determines the spin flow due to the current of polarized electrons.  The magnetizations $\mathbf{M}_L$ and $\mathbf{M}_R$ of the left and right ferromagnets are supposed to lie in the $xy$ plane. They are defined by absolute value $M_L, M_R$ and angle of rotation $\theta, \psi$ as:
\begin{eqnarray} \label{Eq2_94}
\mathbf{M}_L &=& M_L \left(cos \theta, sin \theta, 0\right), \\ \label{Eq2_95}
\mathbf{M}_R &=& M_R \left(cos \psi, sin \psi, 0\right).
\end{eqnarray}
The equations (\ref{Eq2_90})-(\ref{Eq2_93}) are valid in energy range $\epsilon \in \left(\epsilon_F - \frac{V}{2}, \epsilon_F + \frac{V}{2}\right)$, $V$ is the applied voltage, where there are electrons in the left ferromagnet (source) and there are no electrons in the right ferromagnet (sink). Note that only the electrons in this energy range contribute to the electric current in the channel.

The electric and spin current density in the channel may be found as
\begin{eqnarray} \label{Eq2_101}
j_x &=& \frac{e \tau_p}{\hbar} v \frac{\partial \overline{n}}{\partial x}, \\ \label{Eq2_102}
\mathbf{J}^s_x &=& -\frac{1}{2} v \tau_p \frac{\partial \overline{\mathbf{m}}}{\partial x}.
\end{eqnarray}
The boundary conditions are obtained by equating (\ref{Eq2_101}), (\ref{Eq2_102}) at $x = -d$ to (\ref{Eq2_90}), (\ref{Eq2_91}) and (\ref{Eq2_101}), (\ref{Eq2_102}) at $x = 0$ to (\ref{Eq2_92}), (\ref{Eq2_93}). We solve these equations in linear order in $G$ and the product $PP'$.

The total electric current $j_x^\Sigma$ is obtained by integrating the electron current determined by (\ref{Eq2_101}) over the whole energy range:
\begin{equation} \label{Eq2_103}
j_x^\Sigma = \int_{\epsilon_F - \frac{V}{2}}^{\epsilon_F + \frac{V}{2}}{j_x d \epsilon}.
\end{equation}
We perform integration and keep only linear in $V$ terms.

The described approach and electric and spin current in the boundaries (\ref{Eq2_90})-(\ref{Eq2_93}) are valid only when the electrons flow from the left ferromagnet to the right ferromagnet since they explicitly take into account that there are electrons in energy range $\epsilon \in \left(\epsilon_F - \frac{V}{2}, \epsilon_F + \frac{V}{2}\right)$ in the left ferromagnet and there are no electrons in this energy range in the right one. For the opposite direction of electron flow, different equations for electron and spin current should be written. We do not provide them in the text of the paper; however the results obtained for two directions of current may be found below (see Section~\ref{Sec3}).

\section{\label{Sec3}Results and discussion}
Taking into account that the electrons charge is negative, we have for the case considered in Section~\ref{Sec2} $V = \phi_L - \phi_R < 0$. Then the electric current in the absence of magnetic moments is
\begin{equation} \label{Eq3_1}
j^0 = \frac{G}{4} \epsilon_F V.
\end{equation}
We suppose that the electric current is from $L$ to $R$ hereafter (negative sign of current means different current direction). For the positive and negative $V$ the addition to the electric current due to magnetization of source and sink has the form
\begin{widetext}
\begin{eqnarray} \label{Eq3_2}
\delta j_\pm &=& \frac{G}{2} P P' V \epsilon_F^2 \sqrt{\frac{\tau_s}{\tau_p}} \frac{\hbar}{4} \frac{a}{a^2 + b^2} \left(\frac{M_L^2 + M_R^2}{2} \frac{1}{\tanh\left(a \frac{d}{v_F \sqrt{\tau_s \tau_p}}\right)} \right. \\ \nonumber
&&\left. - M_L M_R \left(
\frac{\cos\left(\theta - \psi\right)\cos\left(b \frac{d}{v_F \sqrt{\tau_s \tau_p}}\right)}{\sinh\left(a \frac{d}{v_F \sqrt{\tau_s \tau_p}}\right)} \mp
\frac{\sin\left(\theta - \psi\right)\sin\left(b \frac{d}{v_F \sqrt{\tau_s \tau_p}}\right)}{\sinh\left(a \frac{d}{v_F \sqrt{\tau_s \tau_p}}\right)}
\right)\right),
\end{eqnarray}
\end{widetext}
where $v_F = \sqrt{\frac{2 \epsilon_F}{m_e}}$ is the electron Fermi velocity.
Equation (\ref{Eq3_2}) contains the corrections to resistance due to spin accumulation that are proportional to $M_L^2$ or $M_R^2$ separately. Besides, there is a cross-term proportional to $M_L M_R$. This term is split into two. The first one is proportional to the cosine of the angle between $\mathbf{M}_L$ and $\mathbf{M}_R$ and was observed in experiment~\cite{Jedema2} in parallel or antiparallel magnetization configuration. It is even with respect to the applied external magnetic field $B$. This term describes the ordinary Hanle precession in lateral spin valves.

The second term is proportional to sine of the angle between $\mathbf{M}_L$ and $\mathbf{M}_R$ and thus is odd in $B$. It may exist only in case of noncollinear $\mathbf{M}_L$ and $\mathbf{M}_R$ vectors and thus noncoplanar $\mathbf{M}_L, \mathbf{M}_R, \mathbf{B}$ triplet. The overall dependence of the addition to current due to magnetization of current source and sink on $B$ is neither even nor odd. It should also be noted that the sign of the term $~\cos\left(\theta - \psi\right)$ changes as the voltage $V$ sign changes, but the sign of the term $~\sin\left(\theta - \psi\right)$ does not change. First, this means that the second term is a diode effect. Second, it corresponds to the symmetry of the system with respect to interchange of ``Left'' and ``Right'' notations.

\begin{figure}[t]
\includegraphics[width=3.2in, keepaspectratio=true]{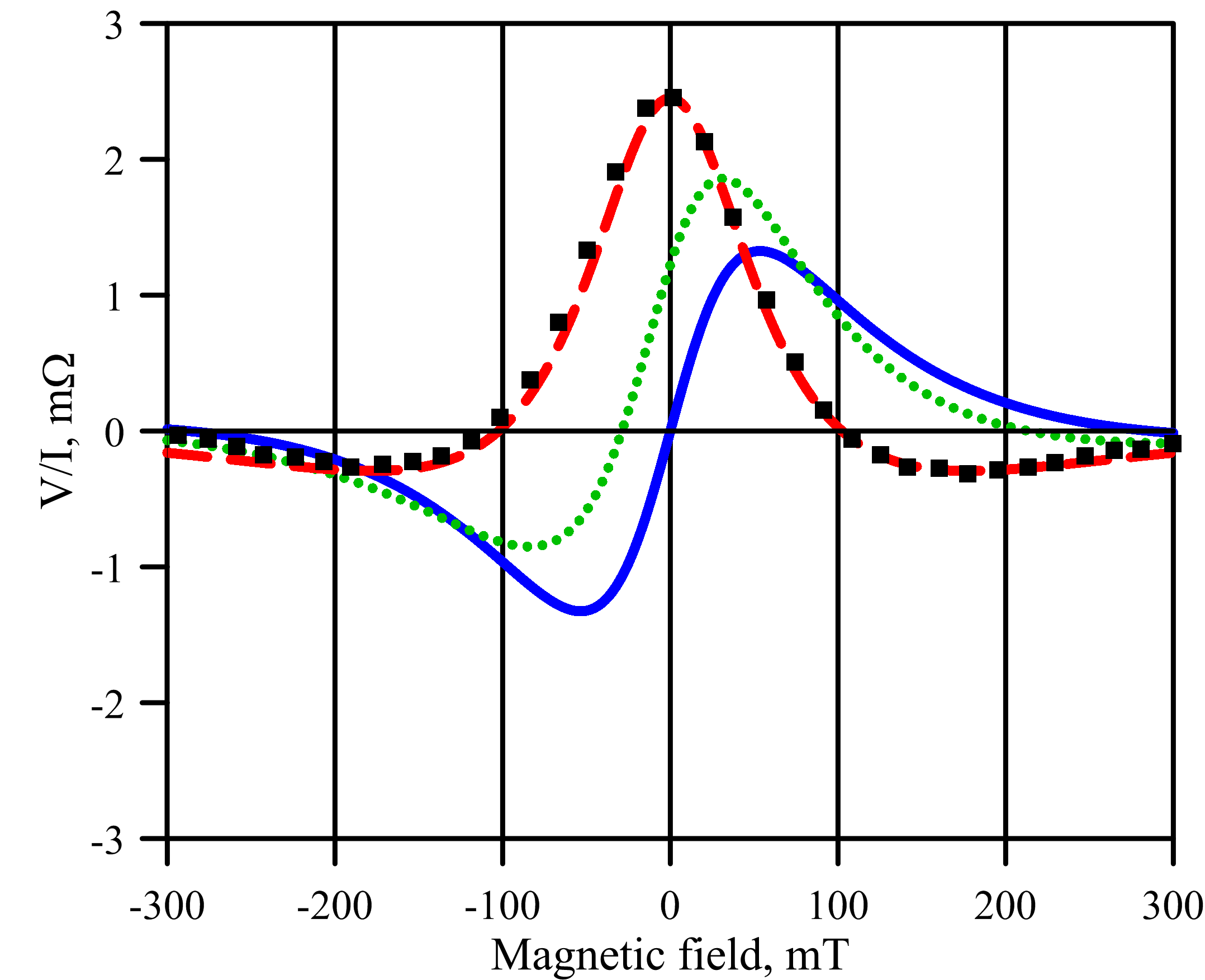}
\caption{\label{Fig2} (Color online) The voltage-to-current ratio dependence on the applied magnetic field for the angle between source and sink magnetization equal to 0 (red dashed curve), $\pi/3$ (green dotted curve), $\pi/2$ (blue curve). The black squares represent experimental data obtained by Jedema et al. \cite{Jedema2}. The red dashed curve is the ordinary Hanle curve.}
\end{figure}

The estimations of the voltage-to-current ratio dependence on the applied field are shown in Figure~\ref{Fig2}. The calculations are done for $d = 1.1$ $\mu m$ aluminium channel at 4.2 K. The amplitude of $\delta j_\pm$ is calibrated by the \emph{V/I} ratio taken from Ref.~\cite{Jedema2}. The fit parameters are the following: $\tau_p\approx2.5\times10^{-15}$ s, $\tau_s\approx90\times10^{-12}$ s and $v_F\approx1\times10^{6}$ m/s. We should note that our parameters lead to diffusion constant $D\approx2.5\times10^{-3} m^2 s^{-1}$ and spin-flip length $\lambda_{SF}\approx480$ nm which are of the same order of value as those taken in Ref.~\cite{Jedema2}.
It is seen that for the collinear case this dependence is even with respect to $B$ and is approximately the same as in the experiment~\cite{Jedema2}. Small difference may be attributed to non-local spin injection which leads to some spread of $d$ and is not taken into account in theory. If $\mathbf{M}_L$ and $\mathbf{M}_R$ are perpendicular to each other this dependence becomes odd. Note that the condition (\ref{Eq2_31}) is satisfied for the parameters of the system for $\tau_e >>4 \times 10^{-10}$ s which conserves the condition $\tau_e >> \tau_s$.

We should note that our model that implies zero temperature works quite well for $T = 4.2 K$. For this temperature and for typical applied voltage of millivolts, the energy value $k_B T$ ($k_B$ is the Boltzmann constant) is less than $e V$. Another important note is that te one-dimensional model works although the channel is much wider than the momentum relaxation length. The spin relaxation length is bigger than the channel width and thus spin diffusion is one-dimensional. The equation for momentum that takes into account perpendicular degrees of freedom may slightly change the fit parameters (such as diffusion constant and spin-flip length) but keep the main effect. Therefore it is reasonable to use such simple model.

According to (\ref{Eq3_2}) the diode effect can be measured by reversing the external magnetic field (similar to nonreciprocal neutron scattering experiments~\cite{Tatarskiy2}). In case of noncollinear ferromagnet electrodes a symmetrical Hanle precession curve will be summed with an odd ``diode'' curve. This effect is of the same order as Hanle effect and undoubtedly can be observed in experiment with a small modification of lateral spin valve structure.

\section{\label{Conclusin}Conclusion}
We have shown that the diode effect in a lateral spin valve with noncollinear magnetizations of two magnetic electrodes is of the same order as the Hanle effect observed in similar systems with collinear magnetizations previously. Our calculations show that the diode effect may be observed in such systems experimentally. So we propose first experiment on electron transport in noncoplanar magnetic device. This would potentially open the way to a new field of noncoplanar spintronics.

\begin{acknowledgments}
This work was supported by RFBR (Grant No.18-32-20036). Discussions with A.A.~Fraerman and O.G.~Udalov are gratefully acknowledged.
\end{acknowledgments}
\nocite{*}
\bibliography{SpinValve}

\end{document}